\documentclass[sigconf]{acmart}

\usepackage{booktabs} 
\usepackage{subfigure}
\usepackage{graphicx}

\setcopyright{rightsretained}

\acmDOI{10.475/123_4}

\acmISBN{123-4567-24-567/08/06}

\acmConference[]{Conference}{Oct. 2017}{Mountain View} 
\acmYear{1997}
\copyrightyear{2017}

\acmPrice{15.00}

\hypersetup{draft} 
\begin{document}
\title{D-Sempre: Learning Deep Semantic-Preserving Embeddings for User interests-Social Contents Modeling}

\author{Shuang Ma}
\affiliation{%
  \institution{The State University of New York at Buffalo}
}
\email{shuangma@buffalo.edu}

\author{Chang Wen Chen}
\affiliation{%
  \institution{The State University of New York at Buffalo}
}
\email{chencw@buffalo.edu}
%
%
%
%
%
%
%

\begin{abstract}
Exponential growth of social media consumption demands effective user interests-social contents modeling for more personalized recommendation and social media summarization.
However, due to the heterogeneous nature of social contents, traditional approaches lack the ability of capturing the hidden semantic correlations across these multi-modal data, which leads to semantic gaps between social content understanding and user interests. 
To effectively bridge the semantic gaps, we propose a novel deep learning framework for user interests-social contents modeling.
We first mine and parse data, i.e. textual content, visual content, social context and social relation, from heterogeneous social media feeds.
Then, we design a two-branch network to map the social contents and users into a same latent space.
Particularly, the network is trained by a large-margin objective that combines a cross-instance distance constraint with a within-instance semantic-preserving constraint in an end-to-end manner. At last, a \textbf{D}eep \textbf{Sem}antic-\textbf{Pr}eserving \textbf{E}mbedding (\textbf{D-Sempre}) is learned, and the ranking results can be given by calculating distances between social contents and users. 
To demonstrate the effectiveness of D-Sempre in user interests-social contents modeling, we construct a Twitter dataset and conduct extensive experiments on it. As a result, D-Sempre effectively integrates the multimodal data from heterogeneous social media feeds and captures the hidden semantic correlations between users' interests and social contents. 

 
\end{abstract}

%
%




\maketitle

\section{Introduction} \label{introduction}


The recent proliferation of social networks has revolutionized everyone's life, and users have an emerging need to explore interesting contents (such as images, videos and articles), which makes the nature of social network gradually shifted from the conventional user-centric networks such as Facebook(i.e. friendship based) to content-centric social networks such as Twitter, Printerest and Flickr. Therefore, user interests-social contents modeling, i.e. capturing correlations between users' interests with the contents on social media, has become a hot topic because of its promising application prospects, e.g. personalized recommendation \cite{Wu:2016:ACMMM,Cai:2015:ACMMM,Tang:2016:AAAI} and summarization \cite{Y.Cong:TIP:2017, Shah:2015:ACMMM, J.Bian:TMM:2015, Schinas:2015:ACMMM, Schinas:2015:ICMR} on social networks.

\begin{figure}[!t]
	\centering
	\includegraphics[scale=0.15]{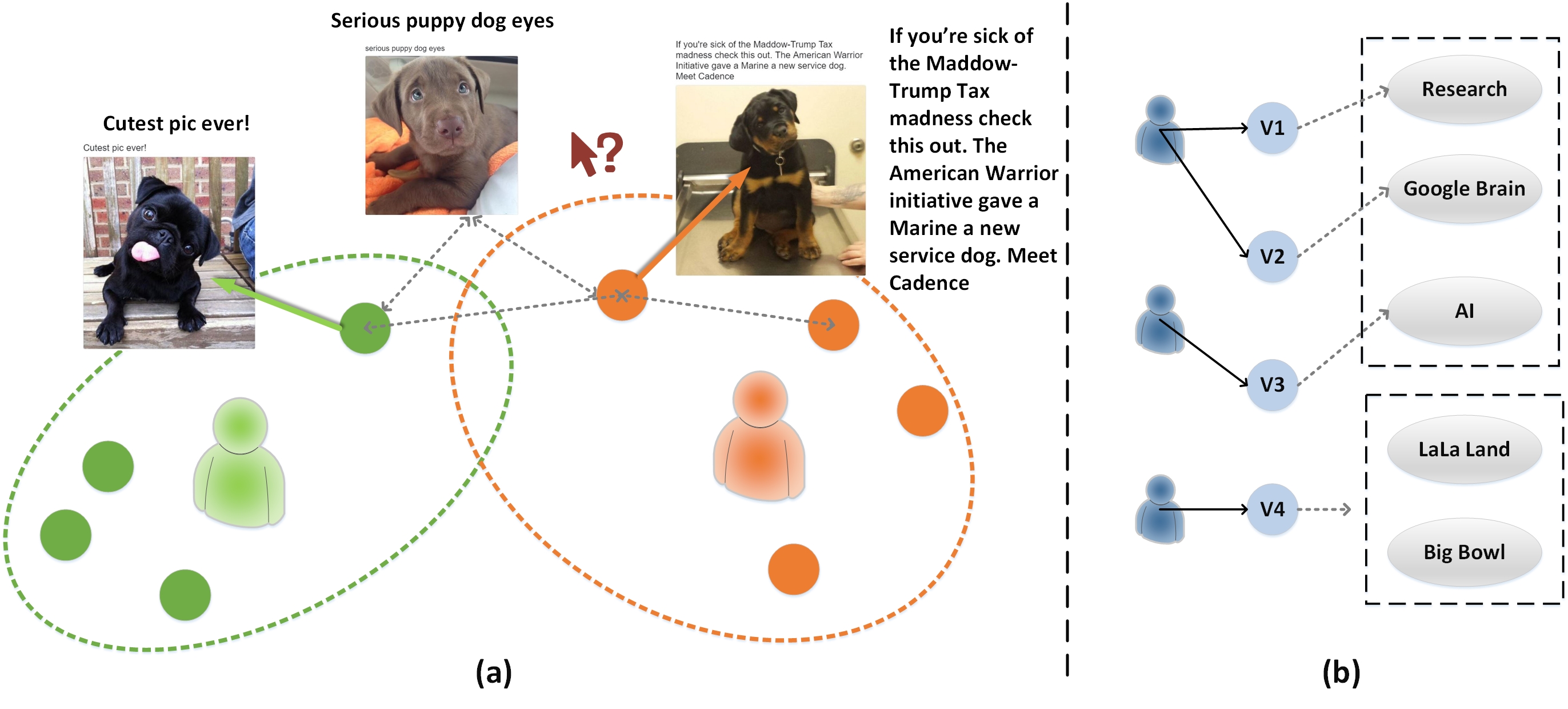}
	\caption{Illustration of the role of hidden semantic correlations in user interests-social contents modeling. (a) shows the data in two semantic spaces (represented by different colors). We pick three posts from Twitter to represent a new query and two samples from different spaces. (b) illustrates a recommender system with three users and four acted posts. The topics in dashed boxes represent the posts' semantic meaning.}
	\label{intro}
	\vspace{-3mm}
\end{figure}

However, the traditional approaches, which primarily focused on structured textual contents are not designed to function effectively in this new era. Because the data from the content-centric social networks have the intrinsic heterogeneous nature, and these traditional methods ignore the rich information in visual contents, social context and social relations. 
Recently, many researchers attempt to adopt more kinds of data sources to improve the user interests-social content modeling, but little progress has been made. This is partially due to the technical difficulties in capturing the hidden semantic correlations from the heterogeneous social media feeds. It has been demonstrated that, the hidden semantic correlations are critical in bridging the social content between the users' interests.
For example, in Figure \ref{intro}(a), we use different colors to represent different users, and the regions within the eclipses represent the hidden semantic spaces. It can be seen that, given a new post, it is hard to determine which semantic space it should belong to. Because without knowing each post's semantic correlation with a specific user, we cannot discriminate them by just looking at the contents (textual contents and visual contents). 
Figure \ref{intro}(b) depicts an example of a recommender system. In this example, user u1 acted on post v1 and v2, user u2 acted v3. Thus, user u1 and u2 have no commonly acted posts. This situation is a typical data sparsity problem (i.e., the known user-post actions are rare comparing with all the user-post pairs). In this situation, the most widely used approaches, i.e. Content-based filtering (CBF) \cite{Saveski:2014:RecSys, Balabanovic:1997:ACM} and Collaborative Filtering (CF)\cite{Su:2009:ACM}, cannot essentially recommend v1 to user u2. Because without knowing the hidden semantic meaning of each post (topics in the dashed boxes), these approaches cannot capture any relation between the posts with users. However, a good recommender system should recommend v3 to user u1 and recommend v1, v2 to user u2, because in this example, user u1 and u2 are probably researchers on AI areas, and post v1, v2, and v3 are probably related to AI.

Driven by this important issue, we present in this paper a novel deep learning framework for the purpose of capturing hidden semantic correlations to effectively bridge users' interests with social contents, and enhance the modeling performance. Specifically, we learn hybrid embeddings by a dedicated two-branch neural network from end-to-end. These hybrid embeddings are directly comparable to rank the social media feeds based on users' interests.
To effectively capture the hidden semantic correlations is indeed very challenging. Due to the heterogeneous nature of social media feeds, data lie in different spaces, which are non-trivial to integrate. 
Some recent approaches \cite{Wu:2016:ACMMM,Chen:2016:ACMMM} attempt to process these multi-modal data, but they often perform separate processing and then simply combine them, which failed in capturing the correlations within multi-modal data.
Several recent Canonical Correlation Analysis (CCA) \cite{Hardoon:2004:CCA} based methods \cite{Gong:2014:IJCV,Gong:EECV:2014, DBLP:journals/corr/KleinLSW14} tried to find projections that maximize the correlation between projected vectors from multi-views. However, as pointed out in \cite{Ma:icml:2015}, CCA is hard to scale to large amounts of data.
The most recent work in \cite{ChenAd:2016:ACMMM} is the first attempt of using deep neural network to directly take both textual content and images as input for integrate modeling. However, less progress has been made due to the naive combination of deep features and user profiles. 

 
To resolve these challenges, we design a dedicated two-branch neural network that maps the social contents and users into a same latent semantic space. We first mine and parse the heterogeneous social media feeds, i.e. textual content, visual content, social context and social relations. 
Then, a large-margin objective is developed to learn the hybrid embeddings from these multi-modal data collectively. Particularly, in the learned latent space, we want data with similar semantic meaning to be close to each other. So we combine a cross-instance distance constraint with a within-instance semantic-preserving constraint so that the hybrid embeddings (term \textbf{D}eep \textbf{Sem}antic-\textbf{Pr}eserving \textbf{E}mbeddings (\textbf{D-Sempre})) are capable of preserving the hidden semantic structure from the training data.
Subsequently, ranking results can be calculated by the learned D-Sempre when given a user and a social post. Generally, the proposed user interests-social contents modeling scheme can be utilized in many social media application scenarios, especially personalized recommendation and summarization. In this paper, to demonstrate the efficacy of the proposed approach, we apply it to the personalized recommendation in Twitter. 
The main contributions of this proposed approach can be summarized into three-fold:

$\bullet$ We propose a novel deep learning framework that can learn hybrid embeddings from heterogeneous social media feeds (i.e. textual contents, visual contents, social context and social relation) in an end-to-end manner. The proposed learning framework is able to characterize the underlying structure of interrelationship among multiple modal entities simultaneously. This shall open a new avenue of deep learning research on heterogeneous information analyzing, and has potential usage value for multi-modal learning.

$\bullet$ We propose a large-margin objective that combines a within-instance semantic-preserving constraint with a cross-instance distance constraint to effectively preserve the hidden semantic structure of the multi-modal data.
Consequently, it can help deal with the data sparsity and cold-start problem.
Because, although a new item has not been seen before, we can still infer a user's preference on new items based on their hidden semantic meanings. These constraints can also provide a useful regularization term for the cross-view matching task. 

$\bullet$ We point out the importance of visual content, external social context and social relation. Moreover, to validate the effectiveness of the proposed approach, we construct a Twitter dataset, and conduct extensive experiments on it. The experimental results show that the D-Spemre is effective in modeling relationship between social contents and users' interests.

\begin{figure*}
	\centering
	\includegraphics[scale=0.2]{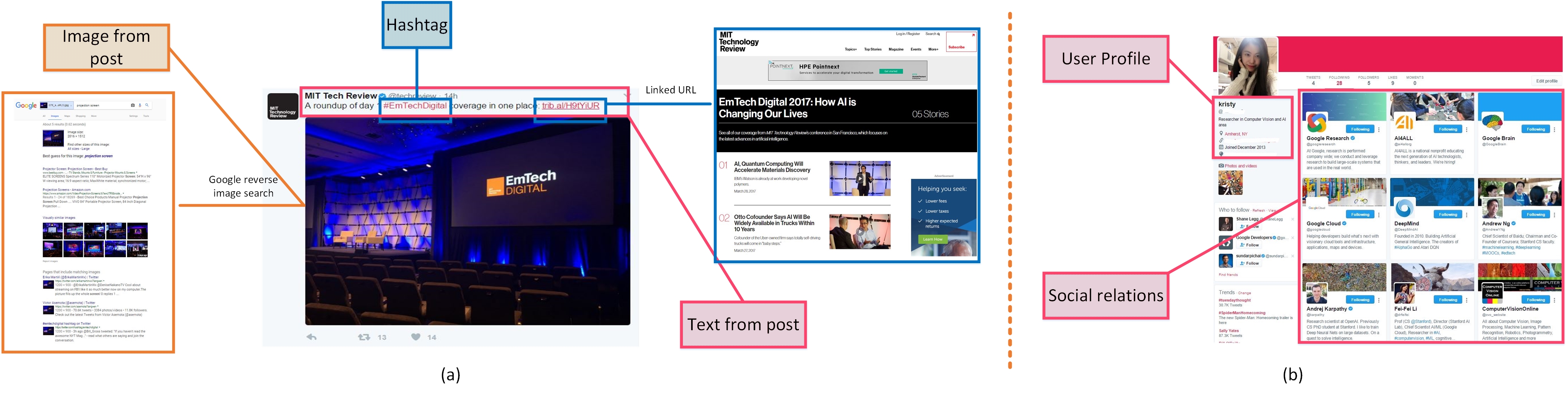}
	\caption{Illustration of the mined data sources. (a) shows the data sources we mined from each social post. (b) illustrates the data sources we mined from a specific query user.}
	\label{data}
	\vspace{-3mm}
\end{figure*}
\vspace{-2mm}
\section{Related Work}


Social network analysis is a classic problem in Artificial Intelligence. Despite the fact that many works have studied user interests-social contents modeling problem, they have primarily focused on the structured textual content such as book, movie, and music \cite{G.Adomavicius:2005:survey}. 
However, they ignore the rich information in visual contents. Instead, they focus solely on modeling users by discovering user profiles and behavior patterns. Visual content is becoming increasingly popular and pervasive in recent years. An image is considered "worth 1,000 words" and it is capable of quickly and directly conveying an idea or describing a scene. Only a few recent work is concentrated on joint modeling of users and visual content for making recommendations \cite{Liu:2015:ACMMM, Liu:2014:ACMMM, Sang:2012:ACMMM}. An interesting technique to automatically select, rank and present social media images was proposed in \cite{McParlane:2014:CIKM}. Sang et al. \cite{Sang:2012:ACMMM} propose a topic sensitive model that concerns user preferences and user-uploaded images to study users' influences in social networks. Liu et al. \cite{Liu:2014:ACMMM} propose to recommend images by voting strategy according to learnt social embedded image representations. 

Typical technologies for the task of user interests-social content modeling include Content-based filtering (CBF) \cite{Saveski:2014:RecSys, Balabanovic:1997:ACM}, Collaborative Filtering (CF)\cite{Su:2009:ACM}, and hybrid of both \cite{Pazzani:1999:CF}. However, it is difficult to directly adopt these technologies for content-centric social media recommendations, due to the heterogeneous nature of social media feeds. To tackle this, some recent works incorporate multi-modal textual contents into collaborative filtering models. Specifically, Chen et al. \cite{Chen:2012:SIGIR} transformed the traditional user-tweet interaction matrix to a user-word matrix before applying matrix factorization. Following
the same idea, Feng et al. \cite{Feng:2013:WSDM} additionally modeled
the user-hashtag interaction, since hashtags are good topic
indicators. Hong et al. \cite{Hong:2013:WSDM} extended the Factorization Machines to jointly model user-tweet relations and the textual tweet generation process. Many efforts have been made to effectively model the users' interests and social contents, but the bottleneck of these purely content based methods comes from the semantic gap between user interests and social content understanding \cite{Jamali:2009:KDD, M.Jiang:2014:TKDE}. More recently, researches have been done to bridging the gap by capturing the social context. For example, Cui et al. \cite{Cui:2014:ACMMM} propose a regularized dual-factor regression method based on matrix factorization to capture the social
attributes for recommendations. In \cite{Wu:2016:ACMMM}, the significance of affective features in shaping users' social media behaviors has been revealed. Similarly, Yang \cite{Yang:2016:AAAI} unveiled how users' emotional statuses influence each other and how users' positions in the social network affect their influential strength about emotion. Chen \cite{Chen:2016:ACMMM} shifted
from the images' pixels to their context and propose a context-aware image tweet-modeling framework to mine and fuse contextual information to model the social media semantics. Moreover, the evolution of social media popularity trends has also been studied and incorporated in generating social media digests \cite{Wu:2016:TMMMM}. However, until now, the existing methods often perform separate processing of user information and images, and then simply combining them. A fully integrated solution needs to be investigated.



\section{Learning D-Sempre}
In this section, we present the proposed D-Sepre learning scheme. We first describe the data mining strategy from heterogeneous data sources in Section \ref{datamining}, and then illustrate the network structure in Section \ref{network}. The training strategy and objective are presented in Section \ref{training}. Since our proposed model learns hybrid representations, it is well suitable for content-centric recommendation tasks. Considering Twitter is the most representative and popular social network, in this paper we will take Twitter recommendation for an example to discuss the utility of our proposed framework. However, in general, the proposed framework can be used for any other social networks, e.g. Flickr, Pinterests, Instagram, etc.

\subsection{Data mining and parsing strategy} \label{datamining}
Given the fact that the social contents come from heterogeneous data sources, how to prepare the training data is not obvious. In this paper, we integrate mine and parse data to effectively represent both social contents and users. Particularly, we utilize both intrinsic contents and extrinsic contexts from each post to generate social content descriptors. Given a query user, personal profiles and social relations are utilized to generate user descriptors. More details will be presented in the following.

\subsubsection{Mining data from social contents}
To effectively generate social content descriptors, we not only mine the intrinsic data, but also the external data of each post that may contain hidden context of the post itself. 
An example is shown in Figure \ref{data}(a), where the intrinsic visual content and textual content are straightforward, i.e. the raw image $I_{En}$ and base text $T_{EN}$ from the post. Particularly, we also collect the hashtags from the base text. It has been demonstrated in \cite{Laniado:2010} that, compared to the textual words of a tweet, hashtags exhibit stronger semantic ties to the post. For example, in Figure \ref{data}(a), the hashtag $\sharp$EmTechDigital describes the topic and event of the post. Therefore, we use hashtag $H_{tag}$ to enhance the intrinsic textual content $T_{En}$. We follow Chen \cite{Chen:2016:ACMMM} to break up hashtags into component words by Microsoft's Word Breaker API \footnote{https://www.microsoft.com/cognitive-services/en-us/web-language-model-api} e.g., $\sharp$EmTechDigital will be broken up as "EM", "Tech" and "Digital". We then combine such component words with base text $T_{En}$ to form the hashtag-enhanced text. 

Despite the intrinsic content, the external hidden context is another important feature. To provide context as well as to circumvent length limitations, Twitter users also embed shortened external links in their tweets. This suggests the external resource is the original context for such tweets, and thus a reliable source for capturing the tweet's semantics. We thus extract the main textual content from the linked URLs. 
We further leverage Google inverse web search engine to obtain the hidden context from from the visual content.
It has been demonstrated in \cite{Chen:2016:ACMMM} that many tweet images originate from news on the Web. To obtain these external contexts, we send each raw image $I_{En}$ as query to Google search engine, then parse the first search engine result page (SERP) to obtain a list of pages that contain the image (including URL and title). We then follow the top ranked links to crawl the actual content of the external page.
Consequently, the information we mined from a post $P$ can be described as follow:

\begin{equation}
{P_{\inf o}} = {I_{En}} + {T_{En}} + {H_{tag}} + Google({I_{En}}) + URL
\end{equation}

\subsubsection{Mining data from users} \label{user_data}
Different from previous works that simply adopt user ID or users' personal profiles to generate user descriptors, e.g. user tags, gender, location and etc. We also consider the social relationship that plays an important role in shaping users' hidden preference features.
As shown in Figure 5(b), given a query user $u$, we mine the user profile (user description and location) and her social relations (all of her followees' profiles). As they are all textual content, we generate descriptors by using the well-known
Word2Vec \footnote{https://code.google.com/archive/p/word2vec/} Specifically, all the mined information are converted to vectors and then vectors are clustered by k-means into 2000 semantic clusters.
Then, the bag-of-words approach works on these clusters. 
The 2000 semantic clusters are pre-computed. In the stage of training and testing, given a new query user, we can directly adopt the bag-of-words approach to generate the 2000-dimension feature vector for user representation.

\begin{figure}[!t]
	\centering
	\includegraphics[scale=0.25]{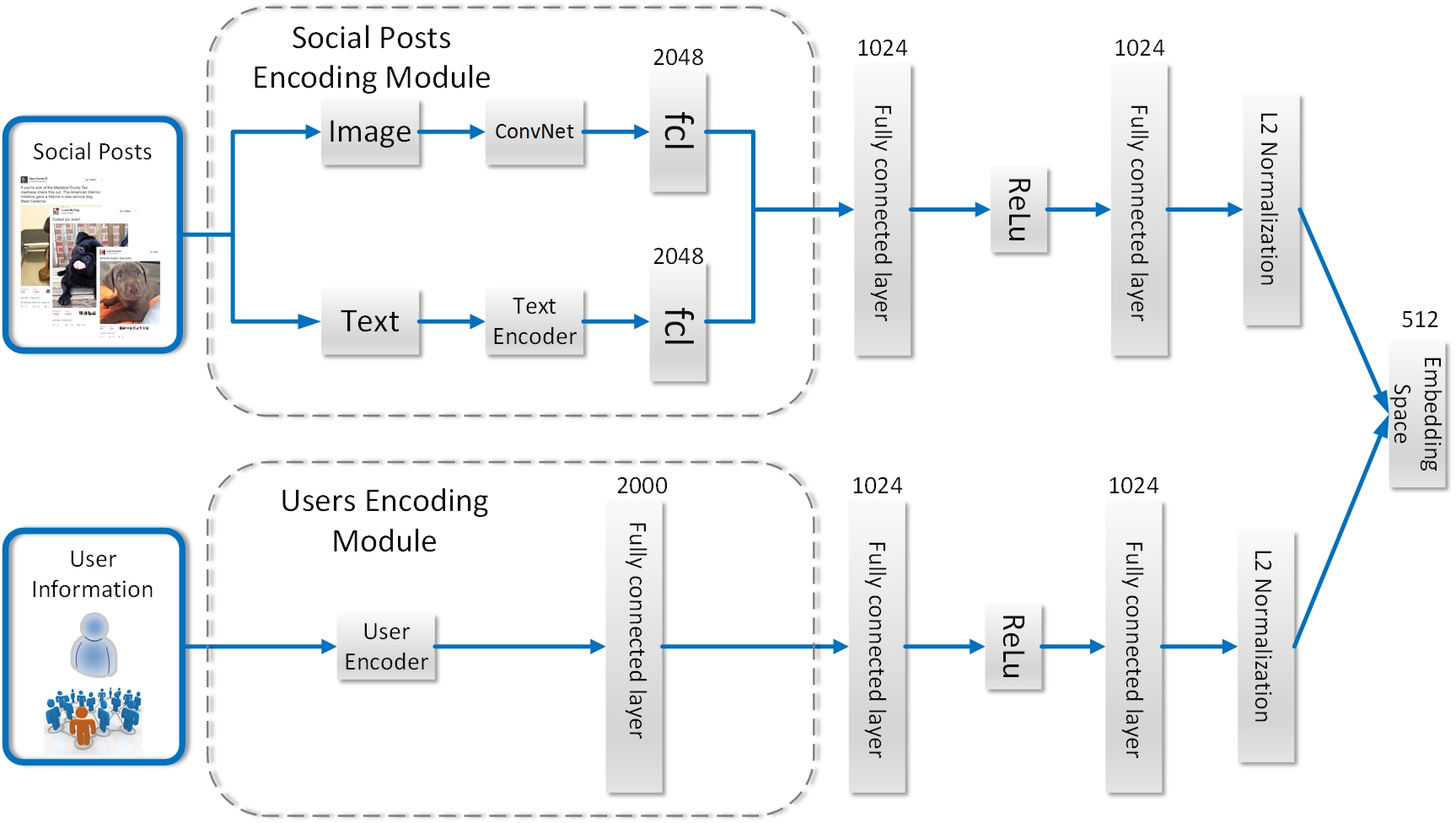}
	\caption{The architecture of the proposed two-branch neural network, where the architecture of ConvNet in the social posts encoding module can be seen in Figure \ref{cnn}.}
	\label{network}
	\vspace{-3mm}
\end{figure}
\subsection{Network Architecture} \label{network}
The architecture of the proposed network is shown in Figure \ref{network}. It can be seen that, the network has two branches. Let $P$ and $U$ denote the collections of the social posts and users. $P$ and $U$ are first feed into the posts encoding module and user encoding module, respectively. Fully connected layers with 1024 hidden nodes are connected on top of each module to generate their own 1024-dimension feature vector descriptors, i.e. $V_u$ and $V_p$. The following two branches consists of fully connected layers with Rectified Linear Unit (ReLU) nonlinearities between them. L2 normalization is followed at the end of each branch. We use the inner product between any two vectors over the learned embedding space to measure similarity, which is equivalent to the Euclidean distance since the outputs of the two embeddings are L2-normalized. In the following, $d(\cdot)$ will denote the Euclidean distance between any two vectors in the embedded space. We train parameters of each of the branch together with the parameters of the embedding layers in an end-to-end manner.

\subsubsection*{\textbf{Social posts and Users encoding modules}}
\begin{figure}[!t]
	\centering
	\includegraphics[scale=0.2]{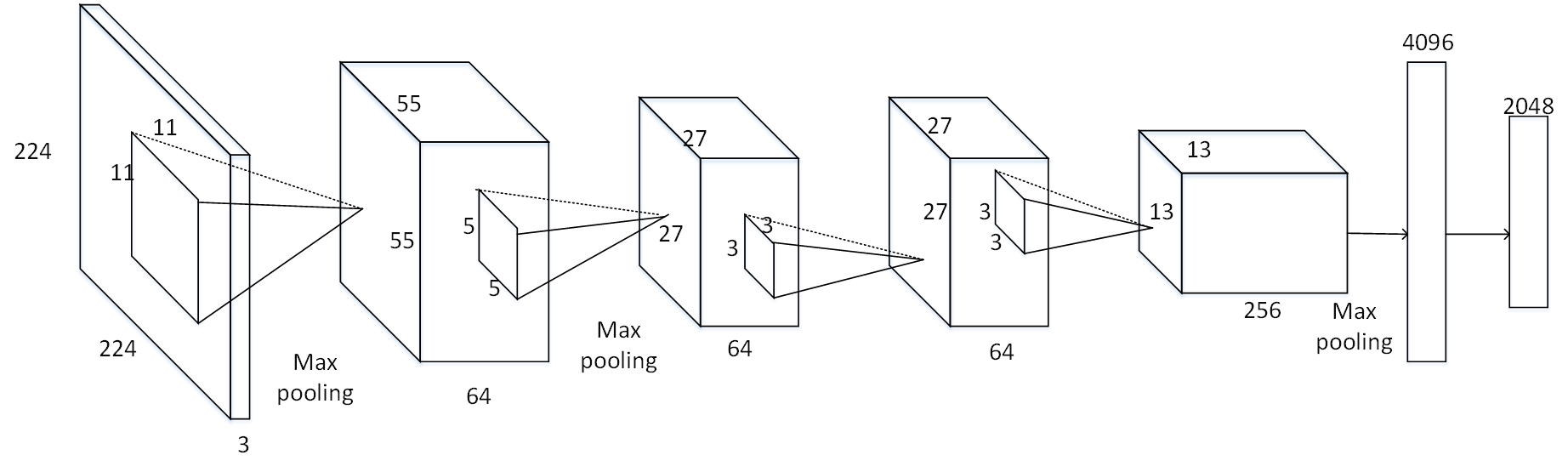}
	\caption{The architecture of the proposed ConvNet.}
	\label{cnn}
	\vspace{-5mm}
\end{figure}

As the data we mined from each social post are consists of visual content (raw image $I_{En}$) and textual content (base text $T_{En}$, hashtag $H_{tag}$ and extrinsic context Google($I_{En}$) and $URL$.) Instead of simply concatenate them together, we design a social posts encoding module to process these two different data sources parallelly, and use a fully connected layer to integrate them together. As illustrated in Figure 2, the social post encoding module consists of two branches. One part, takes raw images as input and follows with a convolutional neural network, which is shown in Figure \ref{cnn}. We do not build a very deep network in consideration of the trade off between performance and training time. The ConvNet consists of 5 convolution layers. It takes 224$\times$224 raw images as input, and outputs 2048-dimensional feature vector for visual content. The second part, takes texts as input and applies Word2Vec text encoding. Same with the first part, a fully connected layer with 2048 hidden nodes is followed to generated a 2048-dimension feature vector for textual content. Subsequently, they are integrated by feeding into a fully-connected layer, and a 1024-dimension feature vector is generated as the post descriptor $V_p$.

To generate the user descriptor is straightforward. Specifically, we first adopt Word2Vec to process data we mined from user domain (user's profiles and social relations) to generate a 2000-dimension feature vector (it has been described in details in Section\ref{user_data}). Then, two fully connected layers are followed to finally output a 1024-dimensional feature vector as the user descriptor $V_u$.

\subsection{Training Objective} \label{training}
Our training objective is a stochastic hinge loss based function that includes an cross-instance distance constraint, together with a within-instance semantic-preserving constraint.

\subsubsection{Cross-instance distance constraint}
Given an user $u_i$, its set of matching samples, i.e. positive sample and negative sample are denoted as ${p_j}^ + $ and ${p_k}^ - $, respectively.
We want the distance between $u_i$ and each positive sample post ${p_j}^+$ to be smaller than the distance between $u_i$ and each negative sample post ${p_k}^-$ by some enforced margin $m$:
\begin{equation}
d({u_i},{p_j}^ + ) + m < d({u_i},{p_k}^ - )\;\;\;\forall \;{p_j}^ +  \in {P_i}^ + ,\;\forall \;{p_k}^ -  \in {P_i}^ - 
\end{equation}

\subsubsection{Within-instance semantic-preserving constraint}

\begin{figure}[!t]
	\centering
	\includegraphics[height=3cm, width=8.5cm]{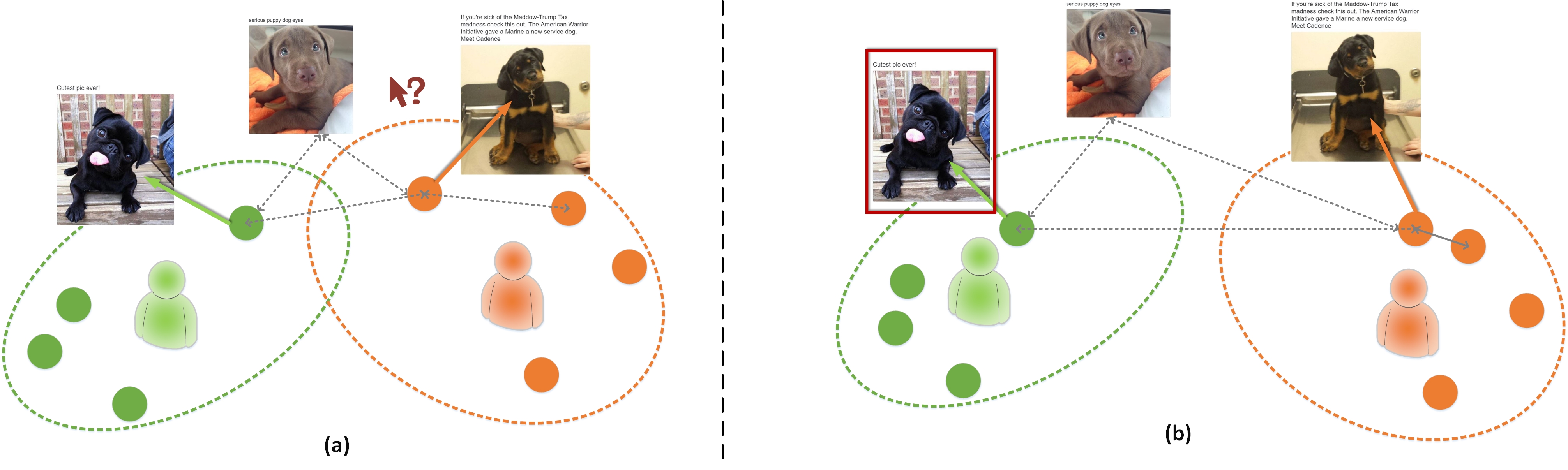}
	\caption{Latent spaces learned by different objectives. (a) shows the latent space that satisfy the cross-instance distance constraint, (b) shows the latent space learned by adding the within-instance semantic-preserving constraint.}
	\label{compare}
	\vspace{-3mm}
\end{figure}

Figure\ref{compare}(a) gives us an intuitive illustration of an embedding space that satisfies the cross-instance matching property. That is, each user is closer to all circles of the same color (representing its corresponding posts) than to any circles of the other color. Similarly, for any circle (posts), the closest user has the same color. However, for the new query, the embedding space still gives an ambiguous matching result since both orange and green circles are very close to it. Driven by this important issue, a question arises: Can we find a way to push semantically similar posts (same color circles) closer to each other, so that the above problem will be mitigated for the new generated embedding? To resolve this technical issues, we add a semantic-preserving constraint to preserve the within-instance structure.

Let $S$ represents a set of semantic clusters, and ${\rm N}{\{ {p_i}\} ^s}$ denotes a set of neighborhood posts of $p_i$ that belong to the same semantic cluster $s$. Then we want to enforce a margin of $m$ between ${\rm N}{\{ {p_i}\} ^s}$ and any point outside of $S$:
\vspace{-2mm}
\begin{equation}
d({p_i},{p_j}) + m < d({p_i},{p_k})\;\;\forall \;{p_j} \in {\rm N}{\{ {p_i}\} ^s}
\end{equation}

In order to generate these semantic clusters, we first adopt Word2Vec to convert all sources of the textual contents (including base text ($T_{En}$), hashtag ($H_{tag}$), as well as textual content from external links ($URL$) and Google search engine ($Google(I_{En})$)) into vectors. Then these vectors are clustered by K-means into 2000 semantic clusters. The 2000 semantic clusters are pre-generated. In the stage of training and testing, given a new query user, we can directly adopting the bag-of-words method to generate the 2000-dimension feature vector. As we can see from Figure\ref{compare}(b), after we add the semantic-preserving constraints, the distances between the new query and the other two samples are much more discriminative in the new generated embedding space.
Consequently, we combine these two constraints into our training objective, the loss function is given by:
\begin{equation}
\begin{array}{l}
L(U,P) = \sum\limits_{i,j,k} {\max [0,m + d({u_i},{p_j}^ + ) - d({u_i},{p_k}^ - )]} \\
\;\;\;\;\;\;\;\;\;\;\;\;\;\; + \lambda \sum\limits_{i,j,k} {\max [0,m + d({p_i},{p_j}) - d({p_i},{p_k})]} 
\end{array}
\end{equation}
where $\lambda$ controls the importance of the semantic-preserving terms. 

Last but not the least question is how to make recommendations for users. This is performed in the following steps. First, the representation of a given query user is generated, it can be calculated and stored in advance, or can be calculated in parallel to accelerate. Then, when feeding a set of posts, the representation of each post can be generated by the proposed social post encoding module. Third, distances are calculated among the set of posts and the user. Finally, K-nearest neighboring posts which having minimum distances are chosen as recommendations.

\section{Experimental Results} \label{experiment}
\subsection{Experiment settings}
\begin{table*}
	\caption{Performance comparisons for textual components}
	\begin{center}	
		\begin{tabular}{c|c|c|c|c|c|c} 
			\hline
			Components & \multicolumn{3}{c}{D-Sempre Precision} & Method & \multicolumn{2}{c}{CITING Precision}\\
			\hline
			&P$@$1&P$@$5&P$@$10 &&P$@$1&P$@$5\\
			\hline
			B:post + user & 0.45 & 0.41 & 0.39&P:post & 0.359 & 0.287\\
			B+Hashtag & 0.47 & 0.42 & 0.39 & P+Hashtag & 0.36 & 0.293 \\
			B+URL & 0.51 & 0.47 & 0.41 & P+URL & 0.381 & 0.300 \\
			B+Google & 0.51 & 0.48 & 0.42 & P+Google & 0.388 & 0.308\\
			Hybrid (textual content) & 0.60 & 0.57 & 0.52 & CITING & 0.419 & 0.319\\	
			\hline
		\end{tabular} 
	\end{center} 	
	\label{textual_components} 
	\vspace{-3mm}
\end{table*}

\begin{table*}
	\caption{Performance comparisons for visual components}
	\begin{center}	
		\begin{tabular}{c|c|c|c|c|c|c} 
			\hline
			Components & \multicolumn{3}{c}{D-Sempre Precision} & Method & \multicolumn{2}{c}{CITING Precision}\\
			\hline
			&P$@$1&P$@$5&P$@$10 &&P$@$1&P$@$5\\
			\hline
			vision      & 0.35 & 0.31 & 0.28 & visual & 0.221 & 0.192\\
			B+vision	& 0.47 & 0.45 & 0.44 & P+Visual	& 0.379 & 0.293 \\
			Hybrid(text+vision) & 0.65 & 0.61 & 0.56 & CITING+visual & 0.425 & 0.313\\	
			Hybrid(text+vision)+semantic & 0.7 & 0.67 & 0.625 & &&\\	
			\hline
		\end{tabular} 
	\end{center} 	
	\label{visual_component} 
	\vspace{-3mm}
\end{table*}
\begin{table}
	\caption{Performance of comparisons for different learning objective and different feature representations}
	\begin{center}	
		\begin{tabular}{c|c|c|c|c} 
			\hline
			&Learning Objective & \multicolumn{3}{c}{D-Sempre Precision} \\
			\hline
			&&P$@$1&P$@$5&P$@$10 \\
			\hline
			Word2Vec & Hybrid & 0.68 & 0.64 & 0.58\\
			& Hybrid + semantic & \textbf{0.7} & \textbf{0.67} & \textbf{0.63}\\			
			\hline
			tf-idf & Hybrid & 0.63 & 0.58 & 0.50\\
			&Hybrid + Semantic & 0.66 & 0.62 & 0.58\\
			\hline
		\end{tabular} 
	\end{center} 	
	\label{tabel1} 
	\vspace{-3mm}
\end{table}

\subsubsection*{\textbf{Dataset}} \label{datasampling}
As there is no publicly available Twitter dataset that covers integrate heterogeneous data sources, we crawl tweets in a user-centric manner by using Twitter API \footnote{https://dev.twitter.com/rest/public} to build up our own dataset . 
We first crawled one week of public timeline tweets (7-14 November 2016) which
resulted in a set of 6,329,306 tweets. From this collection, we randomly sampled about 586 users who had at least 100 followees and 100-3000 followers, and posted at least 500 tweets. These requirements were used to select ordinary but active users, as has been done similarly by \cite{Uysal:2011:UOT}. These 586 users are regarded as target users for our task. We then crawled their latest tweets (up to 3,200 - limited by the Twitter API), personal profile as well as their followee lists, and further crawled the personal profiles of each of their followees. 

In particular, instead of uniform sampling that is widely used for sampling negative instances, we follow \cite{Chen:2016:ACMMM} to adopt the time-aware negative sampling strategy. The main idea is that if a user acted on a post, she/he should also have read other posts that were posted in close temporal proximity to the acted post. Such tweets are then more likely to be true negatives. Specifically, given a positive sample which was retweeted at time $t$
we sample ten non-acted posts in proportion to the time interval $T_r$ centered at $t$, i.e., posts closer to $t$ have a higher chance of being selected. Here we set $T_r$ as 60 min, and this process results in a dataset of 3,692,203 tweets. 
Considering the revolutionary feature on social media, and simulate the real application scenario, we adopt a time-based evaluation. For each user, we use her/his most recent 500 retweets as the test set, with the rest for training. 


\subsubsection*{\textbf{Network Settings}}
In order to accelerate the D-Sempre training process and release the memory burden, we first rain the proposed ConvNet (as shown in Figure \ref{cnn}). Specifically, we adopt ten-way training procedure: the original image is cropped in ten different ways into 224$\times$224 images: the four corners, the center, and their x-axis mirror image. The mean intensity is then subtracted from each color channel. Then, the weights of the ConvNet in the whole network are initialized by the weights of the learned ConvNet.

Our loss involves all triplets consisting of a target instance, a positive match, and a negative match. We sample triplets within each minibatch, because optimizing over all such triplets is computationally infeasible. For the experiments with the within-instance semantic-preserving constraint, we add one more positive sample distinct (belong to different semantic clusters) from the one that already included in the triplets.

We use SGD to train the whole network, and the base learning rate is 0.01, the weight decay is 1e-5 and momentum is 0.9. All the network training and testing are done by using the Caffe deep learning framework \cite{Jia:2014:ACMMM}. 
To accelerate the training and also make gradient updates more stable, we apply batch normalization \cite{Batch:2015:ICML} right after the last linear layer of both network branches. We also use a Dropout layer after ReLU with probability of 0.5. In our experiments, we observe convergence within 60 epochs on average.


\subsubsection*{\textbf{Evaluation Metrics}}
The objective of user interests-social content modeling is to rank the candidate social posts such that the interesting ones are placed at top for the target user. 
To assess ranking quality, we adopt both Precision and Recall at rank k (P@k) as evaluation metrics, which have been widely used for the
recommendation task \cite{Hong:2013:WSDM, Jiao:2010:SIGIR}. For each user, Precision and Recall are defined as follows:
\vspace{-2mm}
\begin{equation}
\Pr ecision@K = \frac{{\# items\;the\;user\;likes\;in\;top\;K}}{K}
\end{equation}

\begin{equation}
{\mathop{\rm Re}\nolimits} call@K = \frac{{\# items\;the\;user\;likes\;in\;top\;K}}{{\# total\;items\;the\;user\;likes}}
\end{equation}
where K is the number of returned items. We compute the
average of all the data' precision and recall in the test
dataset as the final results. 

\subsection{Performance Evaluation}
\begin{figure}
	\centering
	\includegraphics[scale=0.25]{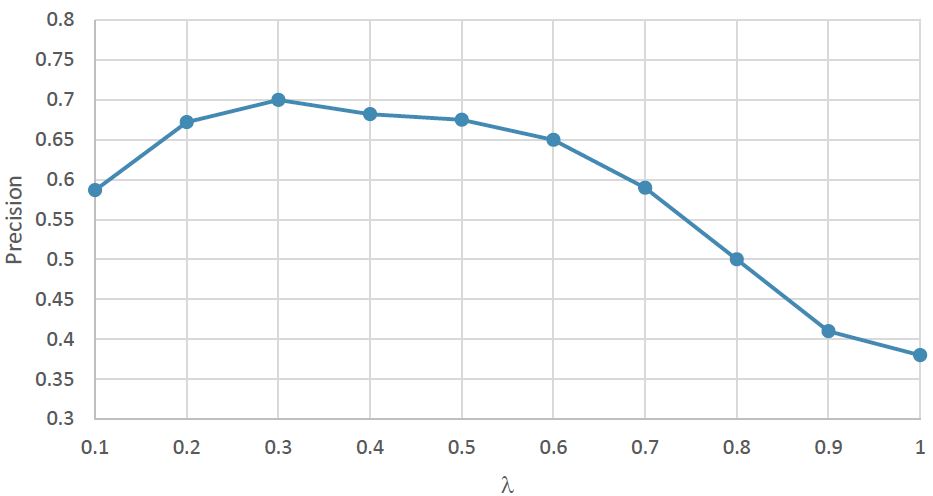}
	\caption{Precision by using different $\lambda$ value.}
	\label{validation}
	\vspace{-3mm}
\end{figure}
\begin{figure*}[!t]
	\centering
	\includegraphics[scale=0.3]{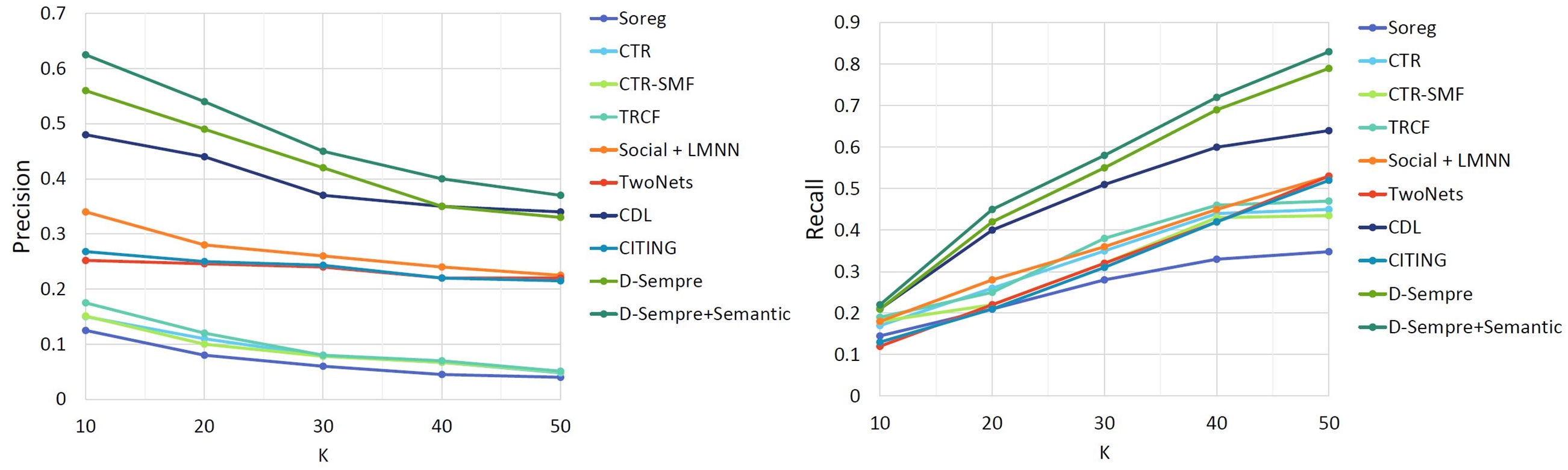}
	\caption{Comparisons of precision and recall with the state-of-the-art.}
	\label{precision}
	\vspace{-3mm}
\end{figure*}

\subsubsection*{\textbf{Evaluation of textual components} }
We first evaluate each textual component to measure their contributions.
To this end, we set the \textbf{Base} as the combination of a tweet's base texts and the user features. We add each component to the \textbf{Base} separately, and assess their performance. For external webpages, we adopt the title and page content (top 20 words in our experiment) in evaluation. 
Table \ref{textual_components} shows the performance of each textual component. In general, all components show a positive impact on the recommendation performance. 

We find that the gains from the two external sources (external URL and
Google Reverse Image Search) are more significant than the internal source (Hashtag). 
This validates the usefulness of external knowledge for interpreting posts' semantics in social network. We further find that the Hashtag component shows minor improvement. The reason might be the low coverage of the data with Hashtags(only 16$\%$ in our dataset). The coverage of URL and Images are 36$\%$ and 92$\%$, respectively. 
Despite that, the gain from URL and Google reverse image search are very close. We find that, in most cases, the top search feedbacks from Google search engine are the images' origin article that exactly points to the same source as the external webpage. 
Thus the data mined from the linked URL and Google search engine are redundant at some extent. This means that only using the feedback information from Google search engine as the feature of visual content is far from enough. We will evaluate the visual component in more details next.

Because Chen \cite{Chen:2016:ACMMM} also adopts these three textual data sources in their proposed \textbf{CITING}, we also make comparisons with their performances. 
For a fair comparison, here we only use textual content as our input.
We can see from table \ref{textual_components} that, our performance by only using Base component is better than the performance of their Base component (post) ( P$@$1: 0.45 vs. 0.359). 
The performance by fusing these three components shows significant improvement (P$@$1: Hybrid(textual content) 0.63 vs. 0.419)
The reasons come from two folds: 1) instead of just using simple index for each user, we adopt the user social relations and user personal profiles to generate user descriptors, which is more effective in shaping users' social behavior. 2) the proposed deep learning framework is much more effective in integrating these multi-modal data and capturing the semantic correlations among these heterogeneous data sources.

\subsubsection*{\textbf{Evaluation of visual components}}
Chen \cite{Chen:2016:ACMMM} claims that the visual content is insufficient in social media interpretation. However, we argue that the visual contents derived from the raw images have significant effects for understanding the social posts' semantic meanings. To validate the usefulness of the visual content, we evaluate the visual components and made comparisons with the performance of CITING \cite{Chen:2016:ACMMM}. The results are shown in Table \ref{visual_component}.
Here we use the raw images (\textbf{vision}) as the visual content baseline. \textbf{B+visual} represents the base text with raw images. 
We can see that, the performance of just using visual content (\textbf{vision}) is even worse than just using \textbf{Base: post+user} component (P$@$1: 0.45 vs. 0.35). It may because the image types on social network are much more diverse than textual contents (for example, there are meme-styled images, text-styled images, and so on ), which make it hard to obtain explicit semantic meanings.  
However, we find that when we combine the visual content with the other components, the performance shows significant improvement (P$@$1: Hybrid(text + visual) 0.65 vs. Hybrid(textual content) 0.6; Hybrid(text+vision) 0.65 vs. vision 0.35), which validate the effectiveness of visual content.
We further add the semantic-preserving constraint, and the result is even better (P$@$1: Hybrid(text+vision) + semantic 0.7 )

However, the performance of CITING with visual content is just 0.425. Upon our deeper analysis, we find that CITING might not adopt an effective way of using visual content. Specifically, they applied GoogLeNet \cite{GoogleNet}, and simply take the top five labels as the description for each image to conduct the experiment. However, 5 category labels are far from enough to represent the visual content. We find that the images from twitter are much more diverse than ImageNet, so simply use the category labels generated by GoogleNet is apparently not appropriate for this specific task, because the classification results is much poorer than that conducted on ImageNet. What's more, the category labels and the best guess from Google search engine are the same in many cases. In this way, using the category labels is just introducing some redundant information, and it will not help to improve the performance. 
Different from CITING, we use the proposed ConvNet to generate visual descriptors, and train our whole network in an end-to-end manner. It can deal with the original visual content and textual content collectively, and capture their correlations directly from the heterogeneous data sources. Therefore, it can be validated that the visual content is a very important component which is under estimated in CITING and many previous works.


\subsubsection*{\textbf{Evaluation of the proposed objective}}
To demonstrate the effectiveness of the proposed large-margin objective, we conduct experiments under different constraints. Specifically, we first train the network by only using the cross-instance distance constraint ($\lambda  = 0$). Then, we combine the proposed within-instance semantic-preserving constraint to compare their results. 
Particularly, to determine the best hyper-parameter in our large-margin objective, we conduct cross-validation, the result is shown in Figure \ref{validation}. We find that the best performance is given when $\lambda=0.3$. 
From Table \ref{tabel1}, we can see that, the performance is much improved by adding the proposed within-instance semantic-preserving constant (both for Word2Vec and tf-idf text representations P$@$1: Word2Vec 0.7 vs. 0.68, tf-idf 0.66 vs. 0.63).

We are also interested in exploring the effectiveness of our approach on top of simpler text representations to validate the effectiveness of the proposed learning scheme. To this end, we pre-process all the sentences with WordNet's lemmatizer \cite{Loper:2002:NNL} and remove stop words, then use tf-idf to encode tweets' texts.
The performance shows better when using the Word2Vec to encode texts. However, the performance for tf-idf encoding method still shows significant improvement compared with CITING \cite{Chen:2016:ACMMM} and many other methods, which will be illustrated in more details in Section \ref{state-of-the-art}.

\begin{figure*}[!t]
	\centering
	\includegraphics[scale=0.25]{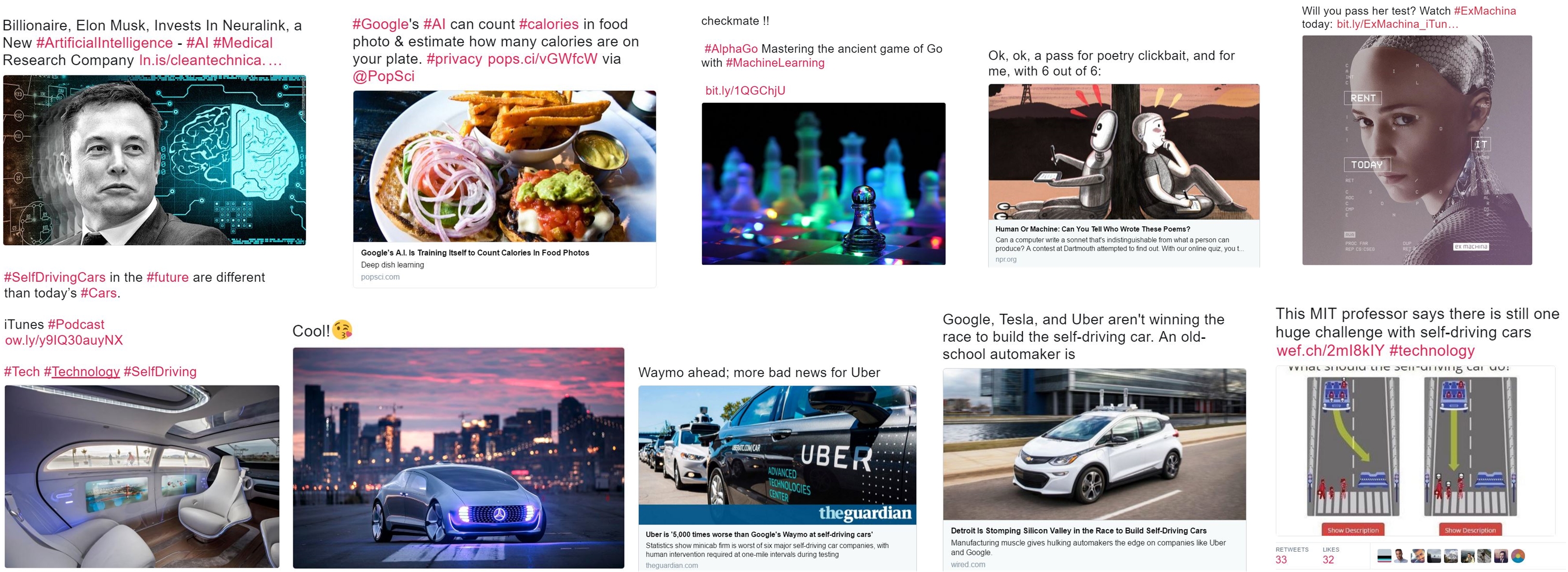}
	\caption{Top five ranked tweets for two users (present in two rows). }
	\label{case_study}
	\vspace{-3mm}
\end{figure*}

\subsection{Comparisons with the state-of-the-art} \label{state-of-the-art}
In this paper, we compare the performance of D-Sempre with the following state-of-the-art approaches:

\textbf{Soreg} \cite{Ma:2011:WSDM}: is a Collaborative Filtering (CF) based approach with social regularization, which uses user-item rating and user social information for recommendation.

\textbf{CTR} \cite{Wang:2011:KDD}: is a topic modeling based CF approach. It uses user-item-tag-rating information to provide an interpretable latent structure for users and items.

\textbf{CTR-SMF} \cite{Sanjay:ICML:2012}: combines user social matrix factorization with CTR. It incorporates user social information additional to the user-item-tag-rating information to automatically infer latent topics and social information to give recommendation.

\textbf{TRCF} \cite{Chen:2016:AAAI}: is a novel CF model, which use topic modeling to mine the semantic information of tags for each user and for each item respectively, and then incorporate the semantic information into matrix
factorization to factorize rating information.

\textbf{CITING} \cite{Chen:2016:ACMMM}: is a context-aware image tweet modelling (CITING) framework that mine and fuse contextual text to model such social media images' semantics. 

\textbf{Social+LMNN} \cite{Liu:2014:ACMMM}: is a metric learning based method that adopts social similarity into a Large Margin Nearest Neighbor (LMNN) regularization for recommendation. 


\textbf{TwoNets} \cite{ChenAd:2016:ACMMM}: is a dual-net deep network, in which the two subnetworks map input images and preferences of users into a same latent semantic space, and then the distances between images and users in the latent space are calculated to make decisions.

\textbf{CDL} \cite{DBLP:CVPR:2016}: is a comparative deep learning (CDL) method, using a pair of images compared against one user to learn the pattern of their relative distances.

The results are shown in Figure \ref{compare}. As we can see that, comparing with the other approaches, the solely textual-content based methods, i.e. Soreg, CTR, CTR-SMF and TRCF, show apparently worse performance in making recommendations. When we add social factors to CTR (i.e. CTR-SMF and TRCF), the performance is improved, which means that the social information is an effective component for user interests-social content modeling.
The approach that adopts hand-craft visual representations, i.e. Social+LMNN leads to better results than the text-based ones, which validates the effectiveness of visual content. CDL shows much better results, almost the third best after our D-Sempre and D-Sempre+semantic, which shows the advantage of deep learning based modeling methods.

Compared to CDL, our approach leads to significant gains for precision and recall. It owes to the superiority of deep network models especially in capturing semantic correlations for the multi-modal data. The carefully designed learning objective which is capable of preserve the inner semantic structure is another reason for the significant improvement. In addition, the intrinsic and extrinsic data sources (i.e., textual content, visual content, social context and social relation) we adopted are also important factors lead to the significant improvement. Note that, CITING also uses these kinds of data sources for recommendation, but it shows poor performance. Thus, how to design an appropriate learning model to effectively utilize these data sources is very important. TwoNets, which adopting deep network model, also has poor performance. It is just better than the text-based methods. It demonstrates that, deep network models do not guarantee great success especially when the task is complicated (learning hybrid representations) and the training data are imperfect (sparse data and implicit feedbacks). The proposed D-Sempre outperforms TwoNets and CDL significantly and consistently, which further demonstrates the effectiveness of the proposed deep learning scheme. 

\subsection{Case study}
It is also instructive to examine individual users and actual
posts to better understand the effectiveness of our proposed approach. To this end, we examine a few users whose recommendations have a large performance gain by using D-Sempre. In Figure \ref{case_study}, we show two typical users and five of their top ranked recommendations in test set. As a consequence, the average recommendation precision of our approach for them are very high (0.83). Taking a closer look at the these tweets, we find many of them trigger multiple data sources, i.e. base text, hashtag, URL and image. 
This validates that, our proposed learning scheme is capable of integrating multi-modal data, and the learned D-Sempre is effective in bridging the semantic gap between social contents and users' interests.
A further investigation shows the effectiveness of external data sources. For example, both the textual content and the visual content from the fourth ranked tweet for user1 are ambiguous for semantic content understanding. Under this circumstance, the external webpage offers more valuable information.
The external data source from Google reverse image search engine also helps in the same way, which can be seen from the second ranked tweet of user2. 
It is also obviously to find that, our proposed approach is capable of capturing the hidden semantic correlations between users' interests with social contents. For example, the top ranked tweets for user1 are relevant to artificial intelligence, and user2 may have more interests in self-driving.

\section{Conclusion}
In this paper, we propose a novel deep learning framework to learn hybrid embeddings, i.e. D-Sempre. The learned D-Sempre can effectively capture the hidden semantic correlations between social media contents and user interests.  
To learn D-Sempre, we propose a dedicated two-branch neural network, and use the proposed large-margin objective to train the multi-modal data (textual content, visual content, and social relation) in an end-to-end manner.
To demonstrate the effectiveness of D-Sempre in user interests-social contents modeling, we construct a Twitter dataset and apply it to the personalized recommendation task. Extensive experiments show that D-Sempre can effectively integrate the multimodal data from heterogeneous social media feeds and captures the semantic correlation between users' interests and social contents.

\bibliographystyle{ACM-Reference-Format}
\bibliography{ref} 

\end{document}